\documentclass[aps,prl,twocolumn,showpacs,preprintnumbers,amssymb]{revtex4}
\usepackage{graphicx}
\usepackage{dcolumn}
\usepackage{amsmath}
\usepackage{amssymb}
\usepackage[pdfauthor={B. Hnat},%
pdftitle={Nonlinear waves in astrophysical plasmas.},colorlinks=true, pdfstartview=FitV, linkcolor=blue, citecolor=blue, urlcolor=blue]{hyperref}
\begin{document}
%\preprint{APS/V2-15/10/02}

\title{Nonlinear waves in the terrestrial quasi-parallel foreshock}
\author{B. Hnat}
\email{B.Hnat@warwick.ac.uk}
\author{D. Y. Kolotkov}
\author{D. O'Connell}
\author{V.M. Nakariakov}
\author{G. Rowlands}
\affiliation{CFSA, Physics Department, University of Warwick, Coventry, CV4 7AL, UK.}

\date{\today}% It is always \today, today,

\begin{abstract}
We study the applicability of the derivative nonlinear Schr\"{o}dinger (DNLS) equation, for the evolution
of high frequency nonlinear waves, observed at the foreshock region of the terrestrial quasi-parallel bow shock.
The use of a pseudo-potential is elucidated and, in particular, the importance of canonical representation in
the correct interpretation of solutions in this formulation is discussed. Numerical solutions of the DNLS equation
are then compared directly with the wave forms observed by Cluster spacecraft. Non harmonic slow variations are
filtered out by applying the empirical mode decomposition. We find large amplitude nonlinear wave trains at
frequencies above the proton cyclotron frequency, followed in time by nearly harmonic low amplitude fluctuations.
The approximate phase speed of these nonlinear waves, indicated by the parameters of numerical solutions, is of the
order of the local Alfv\'{e}n speed.
\end{abstract}
\pacs{Valid PACS appear here}% PACS, the Physics and Astronomy
\maketitle

{\em Introduction. }Magnetised plasmas constitute a dielectric medium in which finite amplitude fluctuations
experience nonlinear evolution and interactions. This can lead to wave steepening \cite{Cohen74}, self-focusing
\cite{Zakharov,Hasegawa} and to parametric instabilities, which may proceed towards fully developed turbulence 
\cite{Robinson}. Understanding the nonlinear evolution of large amplitude Alfv\'{e}n and the fast magnetosonic
waves, their transition to quasi-parallel turbulence as well as their coexistence with the oblique strong turbulence,
is fundamental to the physics of astrophysical collisionless shocks \cite{Kennel88} and solar wind turbulence
\cite{Belcher,Sharma15}. Large amplitude magnetic field fluctuations have been directly observed in the terrestrial
\cite{Eastwood03} and interplanetary \cite{Kajdic12} foreshocks, but are also important for supernova shocks and
the acceleration of galactic cosmic rays \cite{Bell78}. The parallel wave vector component of solar wind turbulence
is poorly understood and its clear detection remains elusive \cite{He11}. 

Various nonlinear equations have been constructed for magnetohydrodynamic (MHD) waves in the past. In a weakly
dispersive medium, nonlinear propagating waves evolve according to the Korteweg-de Vries (KdV) equation and
historically, this model attracted most studies. For example, finite amplitude slow and fast MHD waves
propagating across, or at large oblique angles with respect to the magnetic field, were shown to obey the KdV
equation \cite{Kukutani69}. For MHD fluctuations propagating parallel, or nearly parallel, to the background
equilibrium magnetic field, the derivative nonlinear Schr\"{o}dinger equation (DNLS) proved to be a good
description \cite{Rogister71,Cohen74,Mjolhus89,Kennel88}. The DNLS equation has been studied analytically
\cite{Hada} for homogeneous media and numerically for the inhomogeneous case, e.g \cite{Buti}. A range of
numerical studies was also performed for the cases with dissipation \cite{Ghosh} and heat flux \cite{Medvedev}.
Beyond MHD approximations, high-frequency nonlinear fluctuating structures on the whistler dispersion branch,
exhibiting soliton-like features, have been identified in numerical and analytical studies, e. g.
\cite{Sauer02,Dubinin03}.

Upstream regions of quasi-parallel astrophysical shocks are among the most complex plasma systems. The lack of
collisions requires multi-scale collective dynamics to mediate energy dissipation and isotropy in these regions.
The terrestrial foreshock offers an opportunity to validate predictions of theoretical shock models against
{\em in situ} spacecraft observations \cite{Burgess97,Selzer14}. The presence of nonlinear waves in quasi-parallel
foreshocks has important implications. Wave collapse and self-focusing generate strong electrostatic fields on
kinetic scales, which accelerate particles, modifying their velocity distribution function \cite{Bale98,Graham15}.
Increased viscous and Ohmic dissipation may occur in these regions dominated by strong gradients and shocks.
In a warm plasma and for oblique-propagating waves, the magnetic field fluctuations interact with density and
velocity perturbations and therefore affect the wave propagation speed as well as the wave evolution. Nonlinear
interactions among the waves generate low frequency harmonics (condensate) and these may in turn become unstable
to modulation instability.

We examine the foreshock region of the quasi-parallel terrestrial bow shock and, for the first time, quantify
propagation characteristics and spatial structure of nonlinear high frequency waves by direct comparison of
experimental observations and numerical solutions. First, we discuss the importance of canonical representation
of the first integral of the DNLS in the correct interpretation of the solutions, based on a functional form of the
pseudo-potential. In contrast with previous presentation \cite{Hada} we obtain a pseudo-potential which diverges
to infinity when its argument approaches zero or infinity, a behaviour which guarantees the existence of large
amplitude oscillatory solutions. We then examine observations from the Cluster spacecraft and perform direct
comparison of the nonlinear wave forms found in observations and these obtained from the numerical solutions of
the DNLS.

{\em Exact solutions of the DNLS. }Consider a magnetised plasma with elliptically polarised Alfv\'{e}n waves
propagating in the quasi-parallel {\em z}-direction with the transverse components, $\mathbf{b}=b_x +i b_y$. We define
two characteristic speeds: sound speed, $c_s^2=\gamma k_B T/m_i$ and the Alfv\'{e}n speed $c^2_A=B^2_0/(\mu_0 n_i m_i)$,
where $\gamma$ is the adiabatic index, $T$ is the temperature, $n_i$ is the proton number density and $m_i$ is the
proton mass. The evolution of these left and right-hand polarised modes, is described by the DNLS equation \cite{Hada}:
\begin{equation}
\frac{\partial \mathbf{b}}{\partial t} + \alpha \frac{\partial }{\partial z} \left( \mathbf{b} |\mathbf{b}|^2 \right) - i \mu \frac{\partial^2 \mathbf{b}}{\partial z^2} = 0.
\label{dnls}
\end{equation}
In \eqref{dnls} the temporal and spatial variables have been normalised by the ion gyro-frequency,
$\omega_i \equiv e B_0/m_i$, and the ion inertial length $c_{A}/\omega_i$, respectively.
The constant $\alpha \equiv c^2_{A} / [4(c^2_{A} - c^2_{s})]$,  and $\mu= \pm 1/2$ corresponds to left (-) and
right-hand (+) polarised mode. Writing the transverse components as $\mathbf{b}=b \exp(i \Theta)$, equation
\eqref{dnls} separates into two equations for the fluctuating field amplitude, $b(\phi(t,z))$ and its phase
$\Theta(t, \phi(t,z))$:
\begin{equation}
\frac{\partial b}{\partial t} + 3 \alpha b^2 \frac{\partial b}{\partial z} + \frac{\mu}{b} \frac{\partial }{\partial z} \left(b^2  \frac{\partial \Theta} {\partial z} \right) = 0,
\label{dnls_b}
\end{equation}
\begin{equation}
\frac{\partial \Theta}{\partial t} + \alpha b^2 \frac{\partial \Theta }{\partial z}+ \mu \left( \frac{\partial \Theta}{\partial z} \right)^2 - \frac{\mu}{b} \frac{\partial^2 b}{\partial z^2} = 0.
\label{dnls_P}
\end{equation}
Following \cite{Hada}, the phase angle is assumed to separate into its functional dependence such that
$\Theta(t,\phi)= -\Omega t + P(\phi)$, where $\Omega$ is a constant and $P$ is an {\em a priori} unknown function
of $\phi$.  In the frame of reference travelling with the phase velocity of the perturbation, $V$ (normalised by
the Alfv\'{e}n speed), that is taking $\phi= (1/\mu) (z-Vt)$, which eliminates temporal variations, one finds the
first integral of \eqref{dnls_P} to be $2s dP / d \phi = C + Vs -3s^2$, where $s=\alpha b^2 / 2$ and $C$
is a constant.
 
Equation \eqref{dnls_b}, expressed in the new variable $s$, takes the following form:
\begin{equation}
F(s) \frac{d^2s}{d \phi^2} + \frac{dF(s)}{ds} \left(\frac{ds}{d\phi}\right)^2 = -\sqrt{\frac{2s}{\alpha}} G(s),
\label{dnls_s}
\end{equation}
where $F(s)=1/2s$ and $G(s)=\Omega \mu + V dP/d\phi - 2sdP/d\phi - (dP/d\phi)^2$. The first integral of \eqref{dnls_s}
can then be found using Bernoulli's method, which gives:
\begin{equation}
\frac{1}{2} \left(\frac{1}{\sqrt{2 \alpha s}} \frac{d s}{d \phi} \right)^2 + U_b = C_b,
\label{dnls_sol}
\end{equation}
where 
\begin{equation}
U_b=\frac{1}{4 \alpha}\left[s^3 - 2 Vs^2 + (4 \Omega \mu + V^2 + 2C)s + \frac{C^2}{s}\right],
\label{pot}
\end{equation}
and $C_b$ is a new constant of integration. Treating $s$ and $\phi$ as a generalised coordinate and time of a
pseudo-particle, the first and second terms on the left hand side of \eqref{dnls_sol} represent generalised
kinetic and potential energies written in the canonical form, and the constant of integration $C_b$ can be
interpreted as the total energy of a particle, which depends only on the initial conditions. This subtle point
is important because it is only the canonical functional form of the potential $U_b$ that can be used to predict
the type of solutions of \eqref{dnls} \cite{Dubinov}. If \eqref{dnls_sol} is multiplied by $s$, as was done in
\cite{Hada}, it is no longer in the canonical form. The new potential loses its $1/s$ dependence, and the constant
of motion $C_b$ loses its traditional meaning.
The new potential $U_n=U_b s$ suggests that there is a finite initial energy for which solutions of \eqref{dnls}
are no longer oscillatory, since $U_n$ appears to have a finite value when $s \to 0$.
The coefficients which describe the behaviour of $U_n$ when $s \to 0$ are in this case no longer related only to
the initial condition. In contrast, the form of $U_b$ clearly shows that $U_b \to \infty$ when $s \to 0$ and when
$s \to \infty$, so the solutions of \eqref{dnls} are oscillatory for the arbitrarily large amplitudes. In order to
compare the oscillatory solutions of \eqref{dnls} with the nonlinear wave forms found in the terrestrial foreshock
we will solve \eqref{dnls_s} numerically, for different initial conditions and using experimentally measured plasma
parameters.

{\em Experimental Data and Methodology. }The dataset corresponds to a foreshock crossing on $20/02/2002$ at
$16$:$56$-$17$:$52$UT. We examine three sub-intervals, hereafter referred to as I1, I2 and I3, each a few tens of
seconds long. This foreshock crossing has been studied before in the context of wave characteristics and
temperature anisotropy \cite{Narita03,Selzer14}. We use magnetic field measurement of $\sim \!\! 22.4$ samples per
second from Cluster FGM instrument \cite{Balogh} and $4$ second averaged measurements of plasma parameters from
the CIS-HIA instrument \cite{Reme01}. The transverse magnetic field fluctuations, described by \eqref{dnls}, are
obtained using minimum variance coordinates, which is equivalent to solving an eigen-value problem for the measured
magnetic field variance matrix \cite{Sonnerup98}.
\begin{figure}[h!]
\includegraphics[width=1\columnwidth]{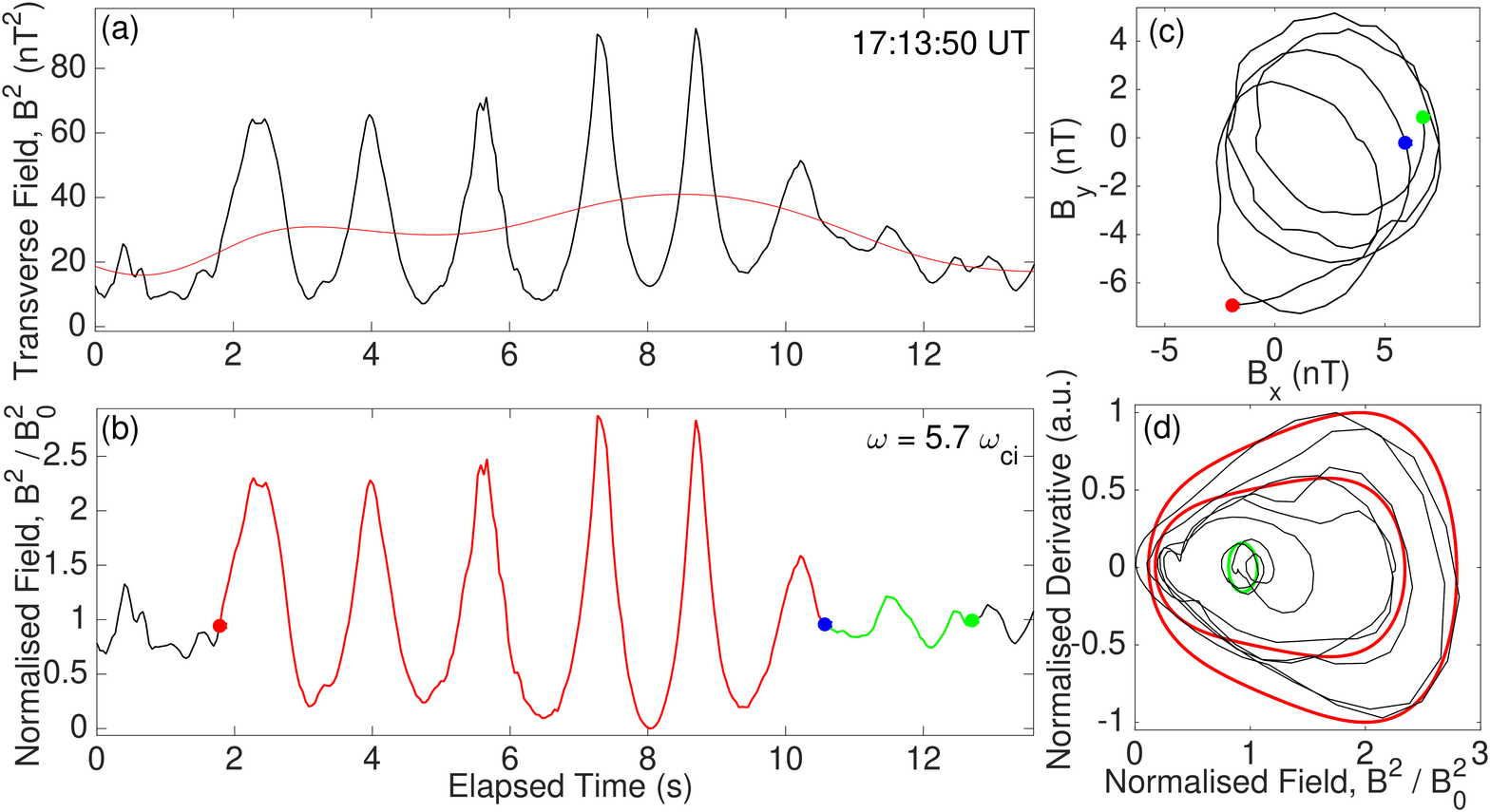}
\caption{Nonlinear wave forms in the magnitude of the transverse magnetic field fluctuations for interval
I1. Panel (a): original squared magnitude of transverse fluctuations (solid black) and a non-harmonic trend
(solid red). Panel (b): normalised and de-trended signal with colour dots marking the start of nonlinear
waves (red), the end of the nonlinear waves/start of nearly harmonic waves (blue) and the end of nearly
harmonic waves (green). The frequency of the red signal, in the spacecraft frame, is given in the top right corner.
Panel (c): hodograms of the transverse components used in (a). Panel (d): the phase space of the signal shown in
(b) (solid black) and the equivalent trajectories obtained from the numerical solutions of \eqref{dnls_s} for
parameters: $C_b=7.84$, $V=4.3$, $C=3.7$ (nonlinear solution, red line) and $C_b=3.91$, $V=8$, $C=7.2$
(linear solution, green line).}
\label{fig1}
\end{figure}

Magnitude of the transverse magnetic field observations is processed using the Hilbert-Huang transform (HHT)
spectral technique \cite{Huang98}, which is suitable for analysing non-stationary and non-harmonic fluctuations.
Unlike traditional spectral methods based on the Fourier decomposition and wavelets, which are restricted
by an a priori assignment of harmonic basis functions, the HHT technique uses the empirical mode decomposition
(EMD), which expands the given signal onto a basis derived directly from the data. The stability of intrinsic
modes detected with the EMD is evaluated using the noise--assisted ensemble empirical mode decomposition (EEMD)
\cite{Wu09}. We note that we do not examine each empirical mode separately, but rather use this technique as
a filter allowing us to find a non-harmonic trend in the signal, which can then be subtracted. This filtered signal
is used to produce phase space portraits, which are compared directly to the ones obtained by numerical integration
of the second order equation \eqref{dnls_s}.
\begin{figure}[h!]
\begin{center}
\includegraphics[width=1\columnwidth]{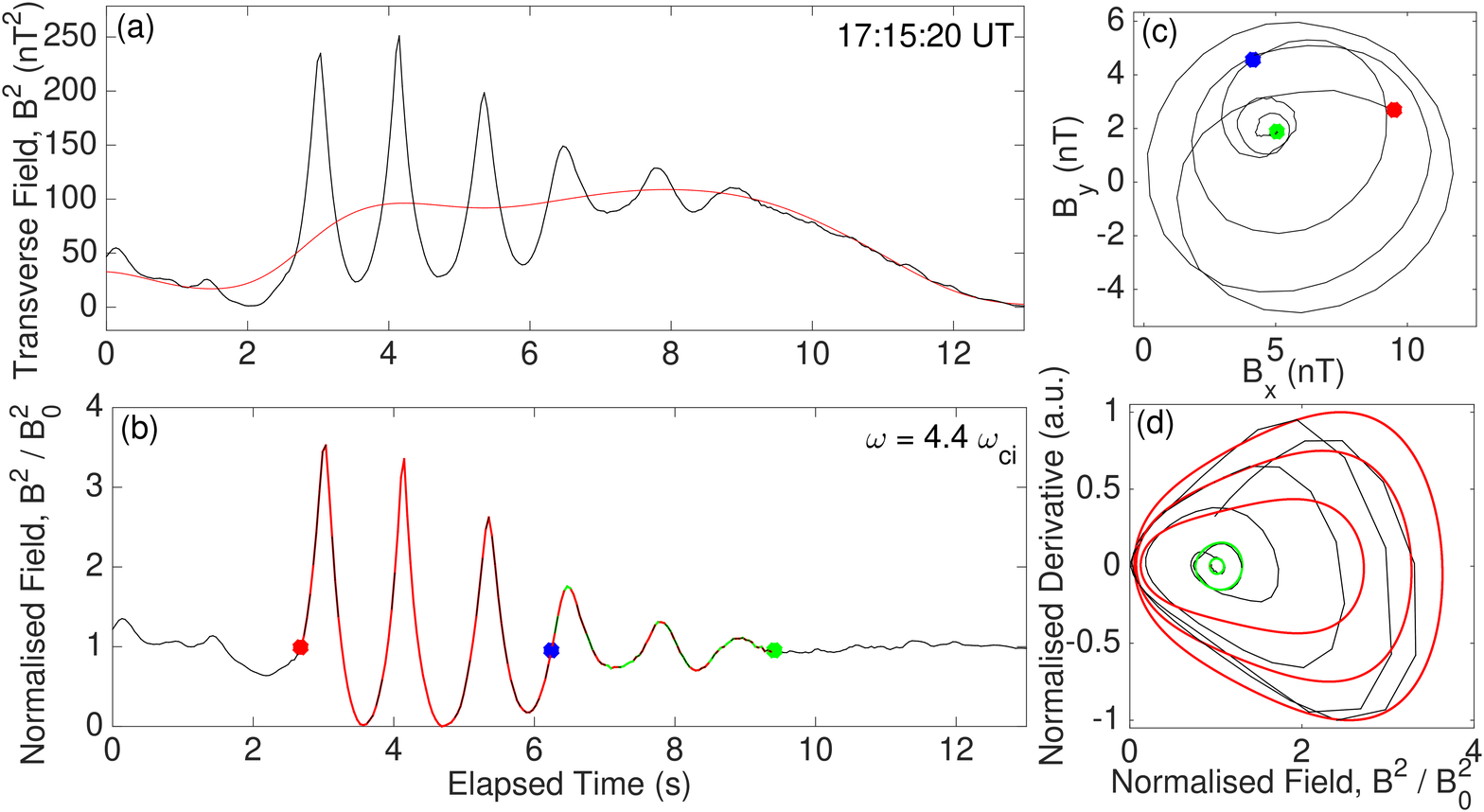}
\caption{Same as figure \ref{fig1} for interval I2. Parameters of numerical solutions: $C_b=25.08$, $V=4.3$, $C=3.7$
(nonlinear solution, red trace) and $C_b=5.31$, $V=8$, $C=7.2$ (linear solution, green trace).}
\label{fig2}
\end{center}
\end{figure}

\begin{figure}[h!]
\begin{center}
\includegraphics[width=1\columnwidth]{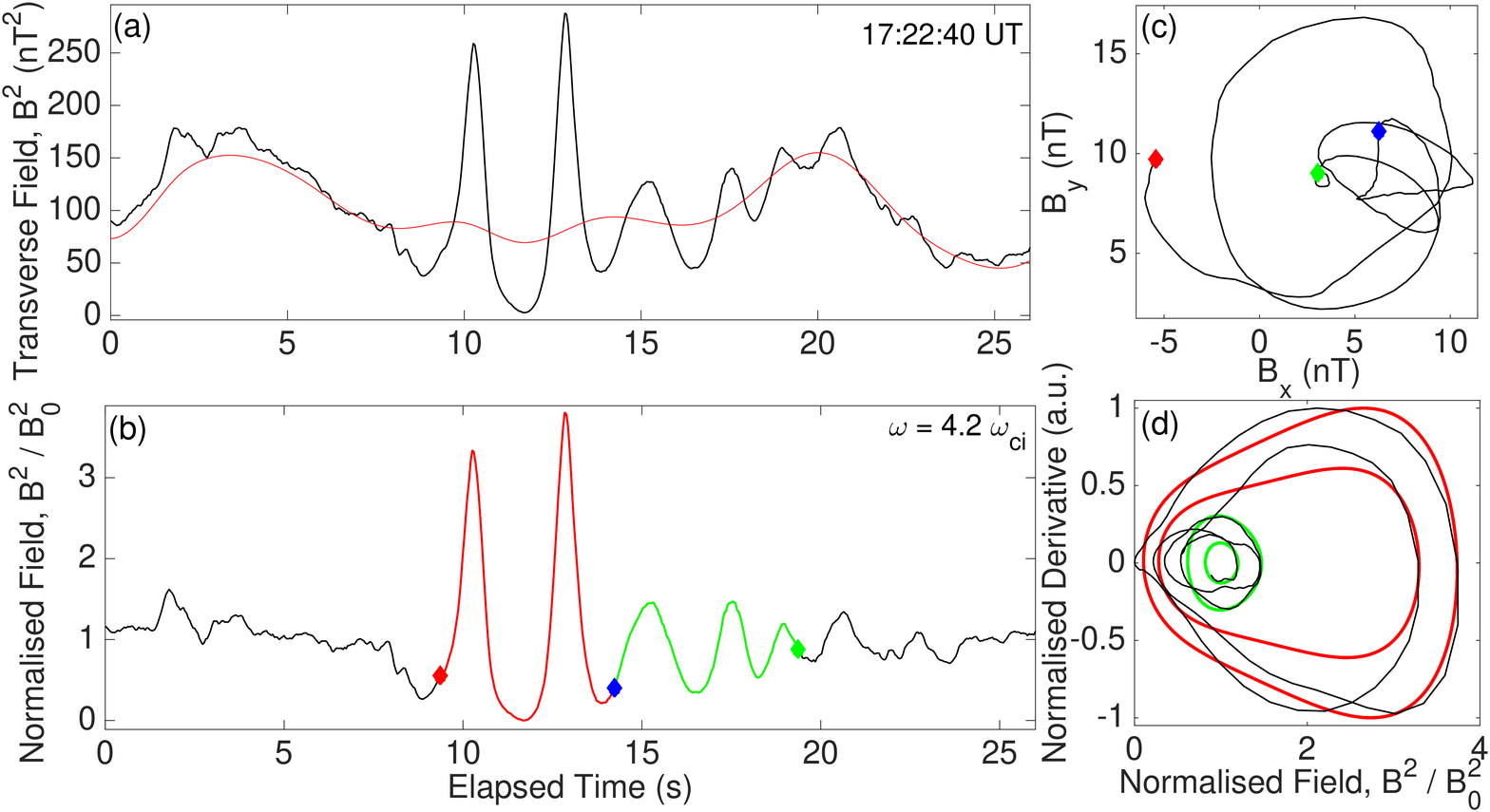}
\caption{Same as figure \ref{fig1} for interval I3. Parameters of numerical solutions: $C_b=16.13$, $V=5.5$, $C=4.8$
(nonlinear solution, red trace) and $C_b=7.33$, $V=9$, $C=8.4$ (linear solution, green trace).}
\label{fig3}
\end{center}
\end{figure}

{\em Results and Discussion. }Figures \ref{fig1}-\ref{fig3} show nonlinear large amplitude quasi-parallel transverse
waves for intervals I1, I2 and I3, respectively. Panel (a) of each figure shows the squared magnitude of the original 
transverse fluctuations, $B^2_t(t)$, in black and a non-harmonic trend, $T(t)$, determined with the EEMD technique,
in red.
We subtract this trend from the original signal and normalise the residue by the mean value of the first and the
last point of the trend. This new signal, $S(t)=2 [B^2_t(t) - T(t)]/[T(1)+T(N)]$, where $N$ is the number of points
in the signal, is plotted in panel (b) of each figure. We determine the frequency in the spacecraft frame, in units
of $\omega_{ci}$, for the nonlinear (red) signal and these are given in the top right-hand corner of each panel (b). 
Colour dots mark the start of the nonlinear waves (red), end of nonlinear waves and transition to nearly harmonic
fluctuations (blue) and the end of nearly harmonic fluctuations (green). Panels (c) show hodograms of the minimum
variance transverse components of the magnetic field, which are left-hand polarised for all intervals, in the spacecraft
frame of reference. Previous studies found modes propagating predominantly away from the bow shock for this foreshock
crossing \cite{Narita03}, which implies reversed sense of polarisation with respect to the plasma frame of reference,
indicating that the observed fluctuations are fast magnetosonic waves. This is consistent with $\mu=1/2$, which we
will use in the numerical solution presented below. Panels (d) show the phase space of the signal in black and the
equivalent trajectories obtained from numerical solutions of \eqref{dnls_s} in red and green.  
First, we comment on the non-harmonic trends identified using the EEMD technique in each example. These have the
variability of tens of seconds, which coincides with the period of the ultra low frequency (ULF) waves
\cite{Nakariakov16} as well as with the typical size of the short large-amplitude magnetic non linear structures
(SLAMS) typically found in the quasi-parallel foreshock \cite{Schwartz91}. We emphasise the importance of the EMD
technique and its ability to extract this non-harmonic trend, which is critical in obtaining stationary nonlinear
wave trains presented here.

The main results of this work are contained in panels (b) and (d) of each figure. In each example, we are able to
identify large amplitude nonlinear wave forms (red lines) which are approximately circularly polarised and span up
to several cycles. These nonlinear fluctuations have a characteristic shape, with round minima and narrowly peaked
maxima, their amplitude is a factor $2\!-\!3$ times larger than the background magnetic field and their periods
(in the spacecraft frame of reference) are shorter than that of the ion cyclotron motion. For all three examples,
the nonlinear wave train is followed by small amplitude nearly harmonic oscillations (green lines). However, since
the solar wind velocity is of the order of $400$km/s, that is much higher than the local Alfv\'{e}n speed, the Taylor's
hypothesis implies that we do not observe the temporal evolution of these waves, rather their spatial variation.

Panels (d) in each figure show a direct comparison of the numerical and experimental phase space trajectories.
Multiple solid red lines, as well as the solid green lines, correspond to numerical solutions with different
initial energies, $C_b$, and with $\mu\!=\!1/2$, $\Omega\!=\!1$ and $\alpha\!=\!1$. The exact values of the initial
energy $C_b$, phase velocity $V$ and constant $C$ are given in the caption of each figure. We note that Cluster
measurements give plasma $\beta \! \approx \!2.5$ for these intervals, which would lead to a negative value of
$\alpha$. However, these measurements are likely contaminated by a dense energetic ion beam reflected from the bow
shock. The NASA OMNI data (http://omniweb.gsfc.nasa.gov), which gives solar wind values time-shifted to the Earth's
bow shock gives plasma $\beta$ in the range $0.75 - 0.9$, which is consistent with the value of $\alpha$ used
in numerical solutions. Clearly, there is a good agreement between experimental and numerical solutions, indicating
that the observed fluctuations are consistent with nonlinear waves, rather than superposition of harmonic signals.
The speed, $V$, used in the numerical solutions is modified by the solar wind speed. Recalling that $V$ is normalised
to the Alfv\'{e}n speed, assuming that the true phase speed of the wave is given by $V_{ph}\!=\!V V_A - U_{sw}$ and
using averaged values (in km/s) of $V_A\!=\!86$, $U_{sw}\!=\!420$ for I1, $V_A\!=\!110$, $U_{sw}\!=\!430$ for I2 and
$V_A\!=\!110$, $U_{sw}\!=\!420$ for I3, we obtain $V^{I1}_{ph}\!=\!-53$ km/s, $V^{I2}_{ph}\!=\!43$ km/s and
$V^{I3}_{ph}\!=\!180$ km/s. These are of order of the local Alfv\'{e}n speed in each case.

\begin{figure}[h!]
\begin{center}
\includegraphics[width=0.325\columnwidth]{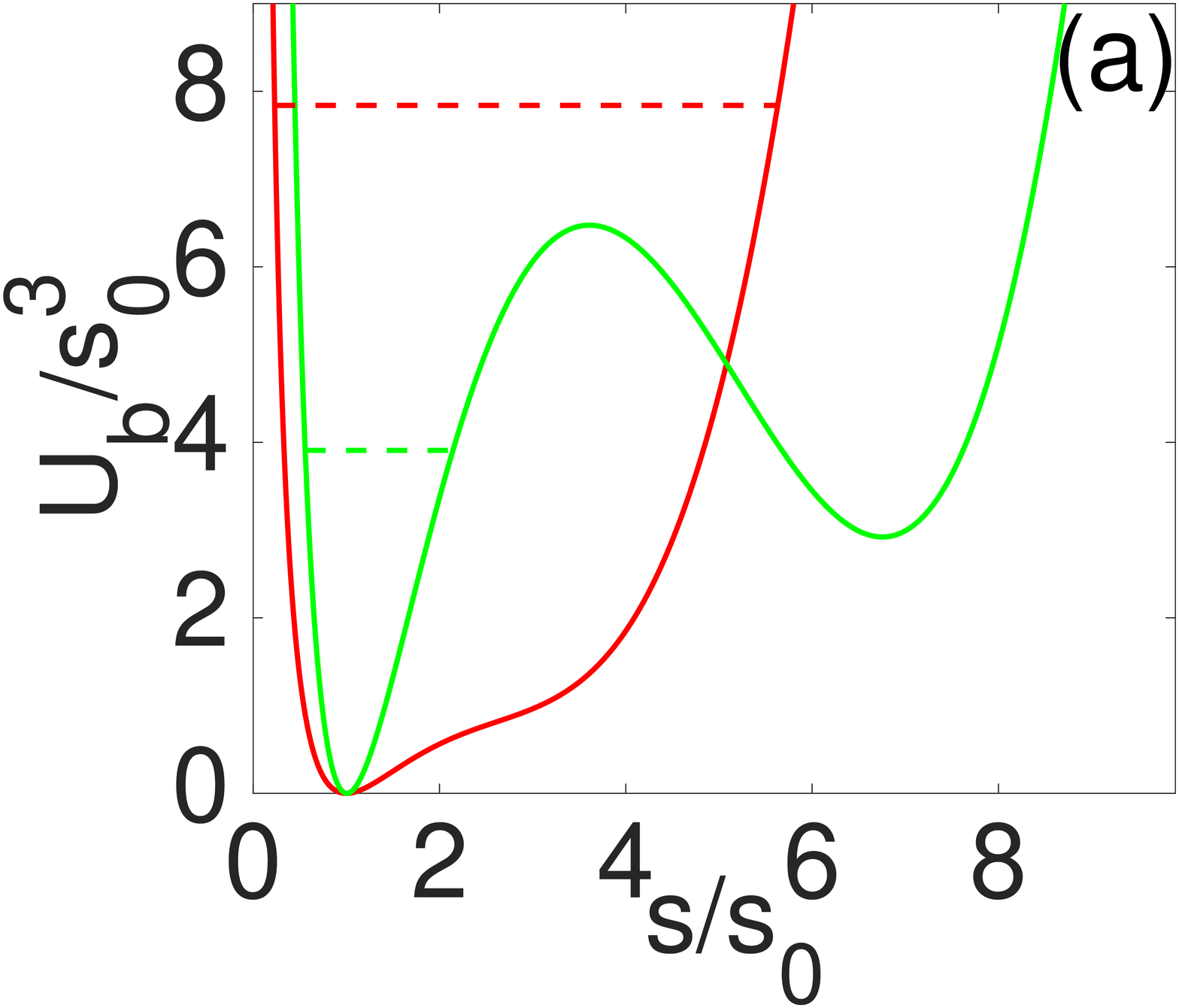}
\includegraphics[width=0.325\columnwidth]{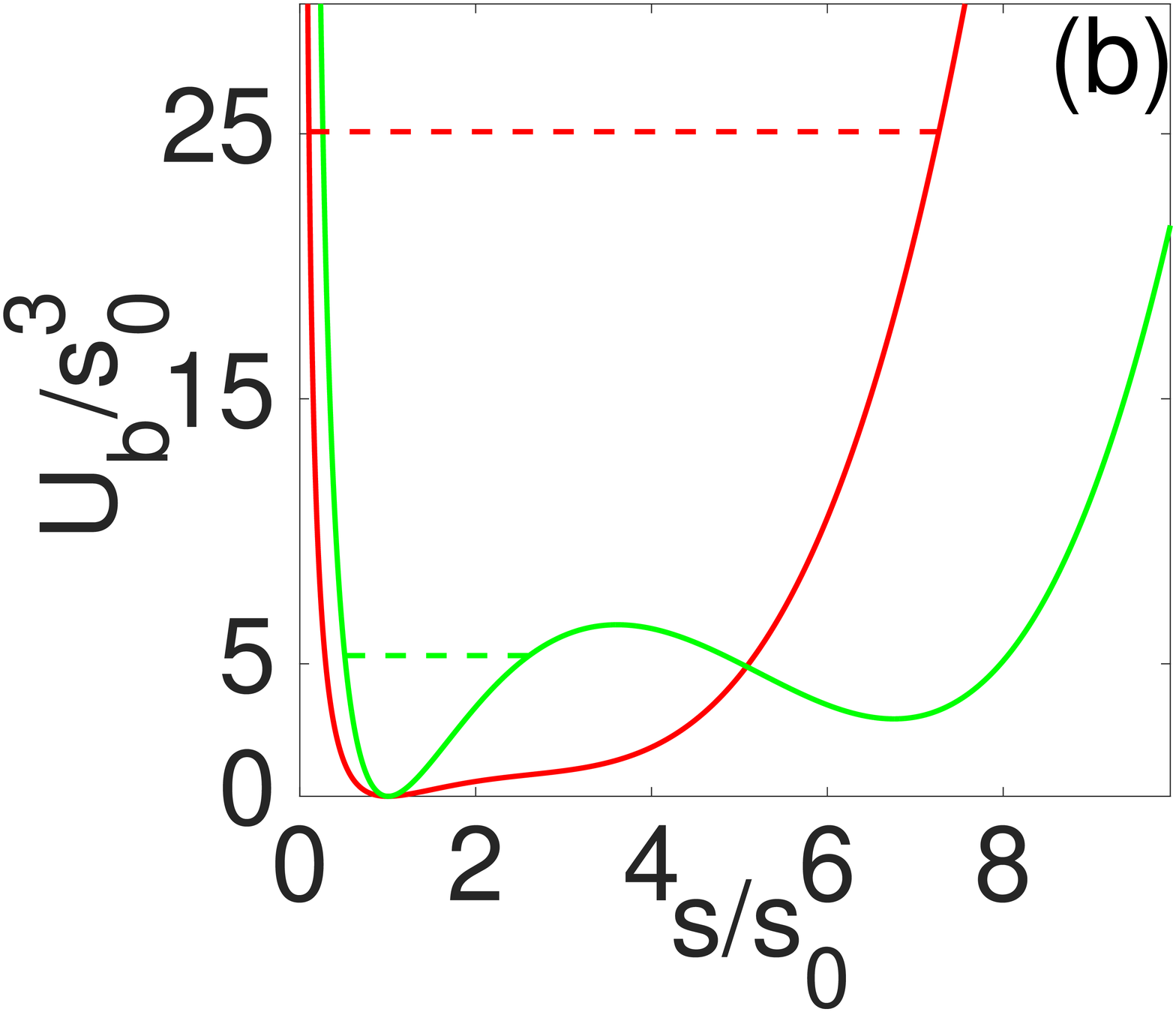}
\includegraphics[width=0.325\columnwidth]{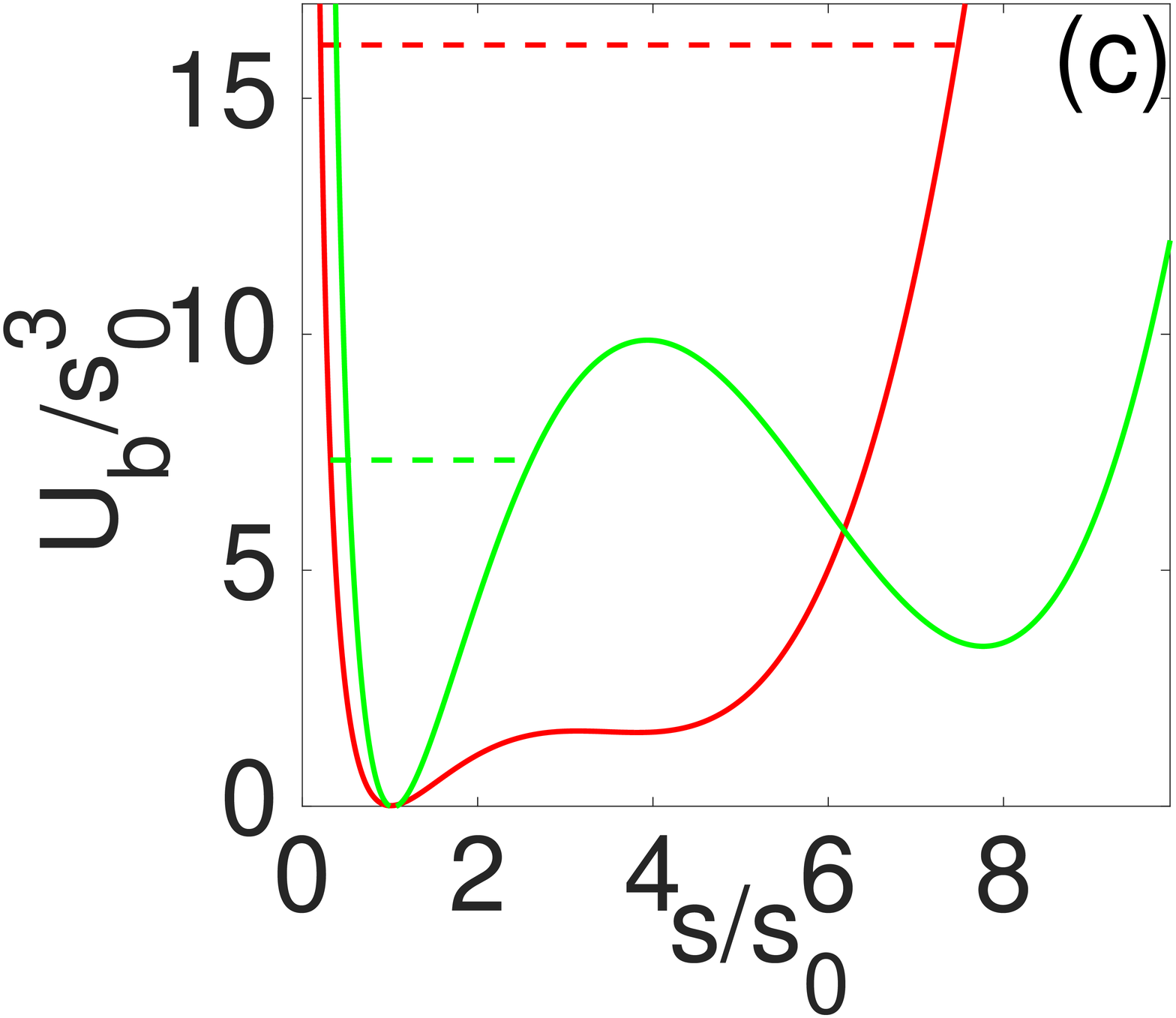}
\caption{The form of potential function $U_b$ in canonical representation for I1 (a), I2 (b) and  I3(c).
In all cases $\mu=1/2$, $\Omega=1$ and $\alpha=1$ and other parameters are these specified in captions of figures
\ref{fig1}-\ref{fig3}. Red and green curves correspond to the outermost nonlinear trajectory and the small amplitude
fluctuations, respectively, as shown in panel (c) of figures \ref{fig1}-\ref{fig3}.}
\label{fig4}
\end{center}
\end{figure}

Figure \ref{fig4} shows normalised functional forms of the pseudo-potential \eqref{pot}, which correspond to
solutions plotted in the phase space panels. Dashed lines in each panel correspond to the initial condition for the
red trajectory and the green trajectories shown in panels (d) of figures \ref{fig1}-\ref{fig3}. The outermost
nonlinear trajectories, plotted as a solid red line in panels (d) of figures \ref{fig1}-\ref{fig3} correspond to
the pseudo-potentials with a single local minimum, which likely are transient states between the single and the
double well potentials. The energy of the observed nonlinear fluctuations are still much higher than any local
extrema or plateau visible in the red curve of figure \ref{fig4}(c). In the classification given in \cite{Hada}
these fluctuations were called algebraic soliton solutions. In contrast, the solid green trajectories of figures
\ref{fig1}-\ref{fig3} represent solutions near the local equilibrium of a double-well potential. The pseudo-potential
reflects the properties of the medium in which the wave propagates. The change of the potential may be interpreted
as the feedback of nonlinear waves on the background plasma. This results in a double-well potential, a solution
corresponding to a large limit in the phase speeds $V$, however the available energy in the system can only sustain
small amplitude oscillations (green trajectories), near one of the equilibrium points.

{\em Conclusions. }We presented the first explicit detection of nonlinear waves with frequencies higher than that
of the ion cyclotron and have shown that these waves are consistent with analytical predictions and numerical
solutions of the DNLS equation. The phase speed of the large amplitude nonlinear waves is approximately equal the
local Alfv\'{e}n speed. Obtained solutions are consistent with parallel propagating fast magnetosonic waves, given
their sense of polarisation and previously obtained direction of propagation \cite{Narita03}. The impact of the
nonlinear waves on the background plasma has been quantified by the change of the pseudo-potential, which shows a
transition from a double well to a single well form. The presence of a double-well potential could, in principle,
support a super-nonlinear wave \cite{Dubinov12}.

As these fluctuations are further advected towards the bow shock, they refract and their greater oblique propagation
angle leads to faster steepening \cite{Hada}. Thus the applicability of the DNLS evolution equation may depend on the
distance from the origin of the fluctuations. Since the quasi-parallel component of plasma turbulence may be treated
as interactions between multiple solitary nonlinear waves, further studies are needed to understand the temporal
evolution of these fluctuations.
%%%%%%%%%%%%%%%%%%%%%%%%%%%%%%%%%%%%%%%%%%%%%%%%%%%%%%%%%%%%%%%%%%%%%%

\end{document}